\documentclass[10pt,twocolumn,letterpaper]{article}

\usepackage{iccv}
\usepackage{times}
\usepackage{epsfig}
\usepackage{graphicx}
\usepackage{amsmath}
\usepackage{amssymb}
\usepackage{graphicx}
\usepackage{amsfonts}
\usepackage{amsmath,bm}
\usepackage{url}
\usepackage{endnotes}
\usepackage{algorithm}
\usepackage{algorithmic}
\usepackage{booktabs}
\usepackage{color}
\usepackage{verbatimbox}
\usepackage{subfigure}
\usepackage{tabularx}
\usepackage{multirow}
\usepackage{verbatim}
\usepackage{pifont}
\usepackage[table,xcdraw]{xcolor}
\usepackage[accsupp]{axessibility}  
\usepackage{footmisc}
\usepackage{balance}
\usepackage[pagebackref=true,breaklinks=true,letterpaper=true,colorlinks,bookmarks=false]{hyperref}

\usepackage{enumitem}

\usepackage{lipsum}

\usepackage[breaklinks=true,bookmarks=false]{hyperref}
\iccvfinalcopy 
\def\iccvPaperID{****} 

\ificcvfinal\fi

\begin{document}


\title{What Matters for Ad-hoc Video Search? A Large-scale Evaluation on TRECVID}

\author{Aozhu Chen\textsuperscript{1*}, Fan Hu \textsuperscript{1*}, Zihan Wang\textsuperscript{1}, Fangming Zhou\textsuperscript{1}, Xirong Li\textsuperscript{1,2$\dagger$}\\
\textsuperscript{1}AIMC Lab, School of Information, Renmin University of China\\
\textsuperscript{2}Key Lab of Data Engineering and Knowledge Engineering, Renmin University of China
}

\maketitle


\begin{abstract}
For quantifying progress in Ad-hoc Video Search (AVS), the annual TRECVID AVS task is an important international evaluation. Solutions submitted by the task participants vary in terms of their choices of cross-modal matching models, visual features and training data. As such, what one may conclude from the evaluation is at a high level that is insufficient to reveal the influence of the individual components. In order to bridge the gap between the current solution-level comparison and the desired component-wise comparison, we propose in this paper a large-scale and systematic evaluation on TRECVID. By selected combinations of state-of-the-art matching models, visual features and (pre-)training data, we construct a set of 25 different solutions and evaluate them on the TRECVID AVS tasks 2016--2020. The presented evaluation helps answer the key question of what matters for AVS. The resultant observations and learned lessons are also instructive for developing novel AVS solutions.
\end{abstract}


\newcommand\blfootnote[1]{%
\begingroup
\renewcommand\thefootnote{}\footnote{#1}%
\addtocounter{footnote}{-1}%
\endgroup
}

\section{Introduction}
\blfootnote{*Aozhu Chen and Fan Hu contributed equally to this work.}
\blfootnote{$\dagger$Corresponding author: Xirong Li (xirong@ruc.edu.cn)}
The core business of Ad-hoc Video Search (AVS) is to develop a video search engine that allows a common user to search for unlabeled short videos by natural-language text. To that end, research efforts are mostly spent on inventing novel cross-modal video-text matching networks including W2VV++ \cite{LiXirong2019W2VVPP}, Dual Encoding \cite{cvpr19-zevr}, SEA \cite{LiXirong2020SEA}, MMT \cite{Gabeur2020MMT}, CLIP4Clip \cite{luo2021clip4clip}, \etc. It can be expected that new models are being developed, which will report even better performance on certain public datasets. Nonetheless, a downside of experimenting on a fully public dataset is that as model developers have complete access to the ground truth of the test set, there is a natural risk of data leakage and thus over-fitting. Therefore, an international benchmark that keeps the ground truth of its test set private (during the campaign phase) is crucial for a real-world assessment of varied AVS models.

Since 2016, the annual TRECVID (TV) evaluation has been such a benchmark for measuring progress on the AVS task \cite{awad2016trecvid}. A specific participant is asked to develop a video retrieval system that retrieves for each test query top 1,000 items from a large collection of unlabeled short videos. The size of the test collection for TV 2016--2018 (comes form IACC~\cite{iacc2009}) is 336k, which increases to one million for TV 2019--2021 (comes from V3C~\cite{v3c2019}). The annual reports by the task organizers provide a good overall picture of the top-performing solutions \cite{awad2016trecvid,awad2017trecvid,awad2018trecvid,awad2019trecvid,awad2020trecvid}. Notice that the solutions often vary in terms of their choices of cross-modal matching models, visual features and training data.  Consider TV20 for instance. The winning solution \cite{zhao2020ruc_aim3} utilizes Dual Encoding as its backbone, but also uses the \emph{ircsn} feature \cite{ircsn} and the VATEX dataset \cite{vatex2019}, both of which are unique when compared to the second-ranked \cite{lirenmin} and third-ranked \cite{wuvireo} solutions. Consequently, what one may conclude from the AVS evaluation is at a high level that does not directly reveal the influence of the individual components. 

Towards filling the gap between the current solution-level evaluation and a desired component-wise evaluation, we propose for the first time a large-scale evaluation based on the TRECVID AVS benchmark series. By training varied state-of-the-art models, features and training data, we build a number of 25 different solutions and evaluate them by answering the test queries from the last five years, from TV 2016 to TV 2020. Such a large-scale evaluation helps answer the key question of what matters for AVS. The resultant observations and learned lessons are also instructive for developing novel AVS solutions.

\section{Proposed Evaluation}


An AVS solution is determined by three key components: cross-modal matching models, visual features, and (pre-)training data. For the sake of reproducibility, we opt for open-source models, visual features extracted by public pre-trained CNNs, and public datasets.

\subsection{Model Zoo}

We select the following five models, among which W2VV++ \cite{LiXirong2019W2VVPP}, Dual Encoding (DE) \cite{Dong2021DE_hybrid} and SEA \cite{LiXirong2020SEA} are used in the previous winning solutions \cite{li2018renmin,zhao2020ruc_aim3} while MMT \cite{Gabeur2020MMT} and CLIP \cite{2021Clip} demonstrate competitive performance on video-text matching tasks. Our model zoo consists of \\
1.  W2VV++\footnote{\url{https://github.com/li-xirong/w2vvpp}}: A cross-modal matching network that encodes a given query by three distinct text encoders, \ie bag-of-words, word2vec and GRU, in parallel. The outputs of the encoders are concatenated into a lengthy vector and projected in a common latent space by an MLP. The given video feature is projected in the latent space by a fully connected (FC) layer. \\
2. DE\footnote{\url{https://github.com/danieljf24/hybrid_space}}: DE uses two multi-level encoding networks of similar architectures, one on the query side and the other on the video side. Different from W2VV++ which accepts only video-level features, DE can handle a sequence of frame-level features. We adopt the latest edition of DE \cite{Dong2021DE_hybrid} which includes a semantic space to improve the model interpretability and matching ability both. \\
3. SEA\footnote{\url{https://github.com/li-xirong/sea}}: SEA exploits multiple text encoders in a multi-space framework, where each encoder corresponds to a specific common space. Consequently, the video-text similarity is obtained by averaging the similarities computed in the individual space. According to \cite{LiXirong2020SEA}, the setup with the following encoders, \ie bag-of-words, word2vec and bi-GRU, performs best for AVS, so we use this setup. \\
4. MMT\footnote{\url{https://github.com/gabeur/mmt}}:  A multi-modal transformer, consisting of a visual transformer to exploit a number of diverse video features and a BERT~\cite{Devlin2019bert} to encode the textual query. \\
5. CLIP\footnote{\url{https://github.com/openai/CLIP}}: Proposed for image-text matching, CLIP consists of a visual Transformer~\cite{dosovitskiy2020image} to extract features from a given image and a textual Transformer ~\cite{radford2019language} to extract features from a given text. By end-to-end training on a web-scale weakly paired multi-modal corpus, CLIP extracts powerful cross-modal features. We use CLIP(ViT-B/32) to extract 512-dim features from video frames and sentences. \\
6. CLIP-FRL: As re-training CLIP is beyond our computational capacity, we instead perform feature re-learning (FRL) \cite{tkde21-frl} on the extracted CLIP features. In particular, a FC layer is used to transform the original 512-dim feature into a new 2,048-dim feature for video-text matching. \\
%
7. SEA-clip: We try to improve SEA by substituting CLIP for bi-GRU. We term the new variant SEA-clip. 

\subsection{Visual Feature Pool}

We collect seven state-of-the-art 2D / 3D deep visual features as follows: \\
1. \emph{rx101}: 2,048-dim frame-level features extracted by ResNeXt-101 trained on the full 21,841-class ImageNet \cite{tomm20-shuffle}. \\
2.  \emph{re152}: 2,048-dim frame-level features extracted by ResNet-152 from the MXNet model zoo\footnote{\url{https://mxnet.apache.org/versions/1.0.0/tutorials/python/predict_image.html}}. \\
3.  \emph{wsl}: 2,048-dim frame-level features extracted by ResNeXt-101,   pre-trained by weakly supervised learning on web images followed by fine-tuning on ImageNet \cite{wsl2018}. \\
 %
4.  \emph{clip}: 512-dim frame-level features extracted by a pre-trained CLIP model (ViT-B/32) \cite{2021Clip}. \\
5.  \emph{c3d}: 2,048-dim segment-level features extracted by C3D~\cite{c3d2015} trained on Kinetics400\footnote{\url{https://github.com/DavideA/c3d-pytorch}}. \\
6. \emph{ircsn}: 2,048-dim segment-level features extracted by irCSN-152~\cite{ircsn} trained by weakly supervised learning on IG-65M\footnote{\url{https://github.com/facebookresearch/VMZ/tree/master/pt}}. \\
7.  \emph{tf}: 768-dim segment-level features extracted by TimeSformer~\cite{timesformer} pre-trained on HowTo100M~\cite{howto100}.

\subsection{(Pre-)Training Datasets}

We collect a number of public datasets that were used by the  participants of the TRECVID AVS task, see Table \ref{tab:datasets}. In particular, the datasets used for training are MSR-VTT \cite{JunXu2016MSRVTT}, TGIF \cite{tgif2015}, and VATEX \cite{vatex2019}. Pre-training datasets include two image collections: MS-COCO \cite{mscoo2014} and GCC \cite{gcc2018}. 


       
\begin{table}[tbh!]
\normalsize
\renewcommand\arraystretch{1}
\centering
\begin{center}
\addvbuffer[3pt 3pt]{
\scalebox{0.75}{
\begin{tabular}{@{}|l|r|r|r|r|r|r|r|@{}}
\hline
\multirow{2}{*}{\textbf{Dataset}}  &\multirow{2}{*}{\textbf{Videos}} &\multirow{2}{*}{\textbf{Frames}} &\multirow{2}{*}{\textbf{Sentences}}   & \multicolumn{2}{c|}{\textbf{Video length (s)}}   \\
\cline{5-6}
& & &&\textit{mean} & \textit{median}  \\\hline
    \multicolumn{6}{|l|}{\textbf{Pre-training}}\\ \hline
        C: MSCOCO & - & 123,287 & 616,767 & - & - \\ \hline
        G: GCC & - & 1,925,125 & 1,925,125 & - & - \\ \hline
        \multicolumn{6}{|l|}{\textbf{Training}}\\ \hline
        M: MSR-VTT & 10,000 & 305,462 & 200,000 & 14.6  & 13.0  \\ \hline
        T: TGIF & 100,855 & 1,045,268 & 124,534 & 18.4  & 3.1  \\ \hline
        V: VATEX & 32,239 & 473,707 & 259,909 & 9.7  & 10.0  \\ \hline

        \multicolumn{6}{|l|}{\textbf{Validation:}}\\ \hline
        TV16-VTT-train &200  &5,941  &400  &5.99  &6.47  \\ \hline
       
        \multicolumn{6}{|l|}{\textbf{Test}}\\ \hline
        IACC.3 (TV16--18) & 335,944 & 3,845,221 & 90 &7.8  &2.2  \\ \hline
        V3C1 (TV19--21) & 1,082,649 & 7,839,450 & 50 & 3.3  & 1.2  \\ \hline
\end{tabular}
}
}
\end{center}

\caption{\textbf{Datasets used in our evaluation}. Frames are obtained by uniform sampling with a fixed time interval of 0.5 second.}
\label{tab:datasets}

\end{table}


\begin{table*}[hbt!]
\normalsize
\renewcommand\arraystretch{1.1}
\centering
\setlength{\belowcaptionskip}{3pt}

\begin{center}
\scalebox{1}{
\begin{tabular}{|r|l|l|l|r|r|r|r|r|r|}
\hline
\multicolumn{1}{|l|}{\textbf{Rank}} & \textbf{Model} & \textbf{Visual features} & \textbf{Training data} & \multicolumn{1}{l|}{\textbf{TV16}} & \multicolumn{1}{l|}{\textbf{TV17}} & \multicolumn{1}{l|}{\textbf{TV18}} & \multicolumn{1}{l|}{\textbf{TV19}} & \multicolumn{1}{l|}{\textbf{TV20}} & \multicolumn{1}{l|}{\textbf{MEAN}} \\ \hline
1 & SEA-clip & \emph{rx101\_wsl\_clip\_\textcolor{blue}{c3d}} & M+T+V-C & \cellcolor[HTML]{FBD0D2}0.223 & \cellcolor[HTML]{F9797B}0.321 & \cellcolor[HTML]{EEF2FA}0.159 & \cellcolor[HTML]{FCE1E4}0.203 & \cellcolor[HTML]{F8696B}0.339 & \cellcolor[HTML]{FBB9BB}0.249 \\ \hline
2 & SEA-clip & \emph{rx101\_wsl\_clip\_\textcolor{blue}{c3d}} & M+T+V & \cellcolor[HTML]{FCE1E4}0.203 & \cellcolor[HTML]{F9797B}0.321 & \cellcolor[HTML]{EAEFF8}0.156 & \cellcolor[HTML]{FCEBEE}0.192 & \cellcolor[HTML]{F97274}0.329 & \cellcolor[HTML]{FBC1C3}0.240 \\ \hline
3 & SEA-clip & \emph{rx101\_wsl\_clip\_\textcolor{blue}{c3d}} & M+T+V-G & \cellcolor[HTML]{FBCDD0}0.226 & \cellcolor[HTML]{FAACAE}0.264 & \cellcolor[HTML]{CDDBEE}0.129 & \cellcolor[HTML]{FCE1E4}0.203 & \cellcolor[HTML]{F97274}0.329 & \cellcolor[HTML]{FBC9CC}0.230 \\ \hline
4 & SEA-clip & \emph{rx101\_wsl\_clip\_\textcolor{blue}{c3d}} & M+T & \cellcolor[HTML]{FCE5E8}0.199 & \cellcolor[HTML]{FBC2C4}0.239 & \cellcolor[HTML]{CDDBEE}0.129 & \cellcolor[HTML]{FCDADD}0.211 & \cellcolor[HTML]{FA9A9C}0.284 & \cellcolor[HTML]{FCD9DC}0.212 \\ \hline
5 & SEA-clip & \emph{rx101\_wsl\_clip\_\textcolor{blue}{tf}} & M+T & \cellcolor[HTML]{FCE5E8}0.199 & \cellcolor[HTML]{FBC6C9}0.234 & \cellcolor[HTML]{D0DDEF}0.131 & \cellcolor[HTML]{FCE0E3}0.204 & \cellcolor[HTML]{FAA8AA}0.268 & \cellcolor[HTML]{FCDEE0}0.207 \\ \hline
6 & SEA-clip & \emph{rx101\_wsl\_clip\_\textcolor{blue}{ircsn}} & M+T & \cellcolor[HTML]{FCE6E8}0.198 & \cellcolor[HTML]{FBC9CB}0.231 & \cellcolor[HTML]{C0D2EA}0.117 & \cellcolor[HTML]{FCE5E8}0.199 & \cellcolor[HTML]{FAACAE}0.264 & \cellcolor[HTML]{FCE2E5}0.202 \\ \hline
7 & SEA-clip & \emph{rx101\_wsl\_clip} & M+T & \cellcolor[HTML]{FCE6E8}0.198 & \cellcolor[HTML]{FBC9CB}0.231 & \cellcolor[HTML]{C3D3EA}0.119 & \cellcolor[HTML]{FCE1E4}0.203 & \cellcolor[HTML]{FBB3B6}0.255 & \cellcolor[HTML]{FCE3E6}0.201 \\ \hline
8 & SEA-clip & \emph{rx101\_re152\_wsl\_clip} & M+T & \cellcolor[HTML]{FCEBEE}0.192 & \cellcolor[HTML]{FBCED1}0.225 & \cellcolor[HTML]{B9CDE7}0.110 & \cellcolor[HTML]{FCE9EC}0.194 & \cellcolor[HTML]{FBB3B6}0.255 & \cellcolor[HTML]{FCE8EB}0.195 \\ \hline
9 & SEA-clip & \emph{rx101\_clip} & M+T & \cellcolor[HTML]{FCE4E7}0.200 & \cellcolor[HTML]{FCE5E8}0.199 & \cellcolor[HTML]{C7D7EC}0.123 & \cellcolor[HTML]{FCE0E3}0.204 & \cellcolor[HTML]{FBBFC1}0.242 & \cellcolor[HTML]{FCE9EC}0.194 \\ \hline
10 & SEA-clip & \emph{wsl\_clip} & M+T & \cellcolor[HTML]{FCF4F7}0.182 & \cellcolor[HTML]{FBD8DA}0.214 & \cellcolor[HTML]{B4C9E5}0.105 & \cellcolor[HTML]{FCFBFE}0.174 & \cellcolor[HTML]{FBC9CC}0.230 & \cellcolor[HTML]{FCF5F8}0.181 \\ \hline
11 & SEA-clip & \emph{rx101\_wsl} & M+T & \cellcolor[HTML]{FCFAFD}0.175 & \cellcolor[HTML]{FCD8DB}0.213 & \cellcolor[HTML]{ACC4E3}0.098 & \cellcolor[HTML]{F9FAFE}0.170 & \cellcolor[HTML]{FBCBCE}0.228 & \cellcolor[HTML]{FCF8FB}0.177 \\ \hline
12 & CLIP-FRL & \emph{clip} & M+T & \cellcolor[HTML]{FCF4F7}0.182 & \cellcolor[HTML]{FCD9DC}0.212 & \cellcolor[HTML]{B6CAE6}0.107 & \cellcolor[HTML]{F8F9FD}0.169 & \cellcolor[HTML]{FCD8DB}0.213 & \cellcolor[HTML]{FCF8FB}0.177 \\ \hline
13 & SEA-clip & \emph{rx101\_re152} & M+T & \cellcolor[HTML]{FCFCFF}0.173 & \cellcolor[HTML]{FCE6E8}0.198 & \cellcolor[HTML]{AEC5E3}0.100 & \cellcolor[HTML]{F8F9FD}0.169 & \cellcolor[HTML]{FCD8DB}0.213 & \cellcolor[HTML]{FAFBFE}0.171 \\ \hline
14 & SEA-clip & \emph{wsl} & M+T & \cellcolor[HTML]{EBF0F9}0.157 & \cellcolor[HTML]{FCE4E7}0.200 & \cellcolor[HTML]{AAC2E2}0.096 & \cellcolor[HTML]{E2E9F5}0.148 & \cellcolor[HTML]{FBCED1}0.225 & \cellcolor[HTML]{F4F6FC}0.165 \\ \hline
15 & SEA-clip & \emph{clip} & M+T & \cellcolor[HTML]{FCF2F5}0.184 & \cellcolor[HTML]{E7EDF7}0.153 & \cellcolor[HTML]{A8C1E1}0.094 & \cellcolor[HTML]{FCFBFE}0.174 & \cellcolor[HTML]{FCDCDF}0.209 & \cellcolor[HTML]{F2F5FB}0.163 \\ \hline
16 & SEA-clip & \emph{rx101} & M+T & \cellcolor[HTML]{F4F6FC}0.165 & \cellcolor[HTML]{FCEFF2}0.187 & \cellcolor[HTML]{AFC6E4}0.101 & \cellcolor[HTML]{EEF2FA}0.159 & \cellcolor[HTML]{FCE6E8}0.198 & \cellcolor[HTML]{F1F4FB}0.162 \\ \hline
17 & SEA & \emph{clip} & M+T & \cellcolor[HTML]{ECF1F9}0.158 & \cellcolor[HTML]{FCE6E9}0.197 & \cellcolor[HTML]{BACDE7}0.111 & \cellcolor[HTML]{E2E9F5}0.148 & \cellcolor[HTML]{FCF5F8}0.181 & \cellcolor[HTML]{EEF2FA}0.159 \\ \hline
18 & CLIP & \emph{clip} & n.a. & \cellcolor[HTML]{FCFCFF}0.173 & \cellcolor[HTML]{FCDDE0}0.208 & \cellcolor[HTML]{A0BBDE}0.087 & \cellcolor[HTML]{D5E0F1}0.136 & \cellcolor[HTML]{F0F3FA}0.161 & \cellcolor[HTML]{E7EDF7}0.153 \\ \hline
19 & SEA-clip & \emph{re152} & M+T & \cellcolor[HTML]{E8EEF8}0.154 & \cellcolor[HTML]{E6ECF7}0.152 & \cellcolor[HTML]{A1BCDF}0.088 & \cellcolor[HTML]{DCE6F4}0.143 & \cellcolor[HTML]{FCE6E8}0.198 & \cellcolor[HTML]{E1E9F5}0.147 \\ \hline
20 & DE & \emph{clip} & M+T & \cellcolor[HTML]{E0E8F5}0.146 & \cellcolor[HTML]{FCF4F7}0.182 & \cellcolor[HTML]{B7CBE6}0.108 & \cellcolor[HTML]{BACDE7}0.111 & \cellcolor[HTML]{D5E0F1}0.136 & \cellcolor[HTML]{D6E1F1}0.137 \\ \hline
21 & SEA-clip & \emph{\textcolor{blue}{tf}} & M+T & \cellcolor[HTML]{C6D6EC}0.122 & \cellcolor[HTML]{D4E0F1}0.135 & \cellcolor[HTML]{8FAFD8}0.071 & \cellcolor[HTML]{BBCEE8}0.112 & \cellcolor[HTML]{E7EDF7}0.153 & \cellcolor[HTML]{D2DEF0}0.133 \\ \hline
22 & W2VV++ & \emph{clip} & M+T & \cellcolor[HTML]{AAC2E2}0.096 & \cellcolor[HTML]{A0BBDE}0.087 & \cellcolor[HTML]{749CCF}0.046 & \cellcolor[HTML]{E1E9F5}0.147 & \cellcolor[HTML]{FCFBFE}0.174 & \cellcolor[HTML]{B9CDE7}0.110 \\ \hline
23 & SEA-clip & \emph{\textcolor{blue}{ircsn}} & M+T & \cellcolor[HTML]{9BB8DD}0.082 & \cellcolor[HTML]{81A5D3}0.058 & \cellcolor[HTML]{81A5D3}0.058 & \cellcolor[HTML]{80A5D3}0.057 & \cellcolor[HTML]{ACC4E3}0.098 & \cellcolor[HTML]{8FAFD8}0.071 \\ \hline
24 & MMT & \emph{clip} & M+T & \cellcolor[HTML]{82A6D4}0.059 & \cellcolor[HTML]{95B3DA}0.076 & \cellcolor[HTML]{6D97CC}0.039 & \cellcolor[HTML]{6994CB}0.035 & \cellcolor[HTML]{769ED0}0.048 & \cellcolor[HTML]{7AA0D1}0.051 \\ \hline
25 & SEA-clip & \emph{\textcolor{blue}{c3d}} & M+T & \cellcolor[HTML]{729BCE}0.044 & \cellcolor[HTML]{7CA2D2}0.053 & \cellcolor[HTML]{5A8AC6}0.021 & \cellcolor[HTML]{6290C9}0.029 & \cellcolor[HTML]{A0BBDE}0.087 & \cellcolor[HTML]{759DCF}0.047 \\ \hline
\end{tabular}
}
\end{center}
\caption{\textbf{Performance of varied combinations of models, features, and training data on the TRECVID AVS tasks}, ranked by their overall performance in descending order. Performance metric: mean infAP.}
\label{tab:infAP_of_each_model_ranked}

\end{table*}

\subsection{Choices of AVS Solutions}
An exhaustive combination of the varied models, features and training data is impractical. 
We have to be selective about the choices of AVS solutions to be evaluated.

First, in order to compare different models, we use \emph{clip} as the fixed feature for its effectiveness and efficiency. Second, for evaluating the influence of the visual features, we use SEA-clip which performs better than alternatives such as SEA, DE and W2VV++ in preliminary experiments. Lastly, when considering the joint use of the multiple (pre-)training datasets, we use SEA-clip and the concatenation of the top-four features, \ie \emph{rx101}, \emph{wsl}, \emph{clip} and \emph{c3d}. 

For the ease of reference, we use M+T to denote the joint use of MSR-VTT and TGIF. Accordingly, M+T+V means the joint use of MSR-VTT, TGIF and VATEX. M+T+V-C indicates the use of MS-COCO for pre-training. The above considerations result in a number of 25 unique solutions in total, reported in Table \ref{tab:infAP_of_each_model_ranked}.

\section{Evaluation Results and Analysis}

The leaderboard is presented in Table \ref{tab:infAP_of_each_model_ranked}. We analyze the results along the following three dimensions, \ie matching models, visual features, and (pre-)training data.

\subsection{Comparing Different Matching Models}

Matching models matters. Given the same feature (\emph{clip}) and training data (M+T), CLIP-FRL (Rank 12) is $2.5$ times higher than MMT (Rank 24). Interestingly, it can be seen from Table \ref{tab:infAP_of_each_model_ranked} that the off-the-shelf CLIP (Rank 18) performs favorably against several skillfully designed models including DE (Rank 20), W2VV++ (Rank 22), and MMT (Rank 24). The result shows the good generalizability of this pre-trained model. Built on the top of this model, CLIP-FRL performs the best when fixing the choice of the visual feature to be \emph{clip}. The peak performance of $0.249$ is reached by SEA-clip (Rank 1), given enhanced visual features and expanded training data. Given these promising results, how to make full use of CLIP in a more computationally affordable manner deserves further investigation.

Per-query analysis on TV20 shows that there remains difficult queries that the current models fail to response, see the bottom rows of Table \ref{tab:per-topic_analysis}. These queries are typically compositive, \eg \emph{two or more people under a tree} (query \#658). The top-performing models return shots that are only partially relevant, \eg trees without people or people on the tree. Understanding why the models fail, \eg whether the queries or the video content are not well represented or both, requires further investigation.


\subsection{Comparing Varied Visual Features}

As Table~\ref{tab:infAP_of_each_model_ranked} shows,  given SEA-clip as a common model and M+T as their training data, solutions using different features have varied performance, ranging from $0.047$ (Rank 25, \emph{c3d}) to $0.212$ (Rank 4). The result suggests that features have a great impact on the performance.
The 2D deep features are in general better than their 3D alternatives. As the videos used for the evaluation are relatively short, see Table \ref{tab:datasets}, modelling the temporal dimension of the video content is not as crucial as the 3D features' original context of human action recognition. Nevertheless, the 3D features remain beneficial when used together with the 2D features. For instance, compared to the solution Rank 7, which uses \emph{rx101\_wsl\_clip}, the solution Rank 4 additionally uses \emph{c3d}, increasing mean infAP from $0.201$ to $0.212$. Note that we use the simple vector concatenation strategy for feature fusion. Hence, we believe there is much room of improvement for feature fusion.


\subsection{Comparing (Pre-)Training Data}
The performance of solutions with different settings of training data is shown in the first four rows of Table~\ref{tab:infAP_of_each_model_ranked}.
The infAP climbed from $0.212$ (Rank 4) to $0.240$ (Rank 2) after adding VATEX to the training data. Comparing the vocabulary of the three training datasets, we find that VATEX has a number of novel keywords  either absent or with very low occurrence in MSR-VTT and TGIF. Consider the TV20 queries for instance. The keyword \emph{sailboats} of query \#644 barely appears in MSR-VTT and TGIF but appears many times in VATEX. Consequently, the solution Rank 4, which is trained on M+T, performs poorly on this query, see Table  \ref{tab:per-topic_analysis}. Meanwhile, CLIP-FRL (Rank 12) uses the pre-trained CLIP (Rank 18) for query embedding, which can effectively handle words it has not seen before. Hence, both CLIP-FRL and CLIP perform reasonably well on query \#644, even though the two solutions do not exploit VATEX. 

The fourth-ranked model is SEA-clip, which already uses CLIP as one of its three text encoders. We analyzed the model behavior, and found out that the number of training epochs SEA-clip took is much larger than that required by CLIP-FRL. As the training process becomes longer, SEA-clip concentrates more on its current training data, \ie M+T, with the capability of the pre-trained CLIP encoder diminished. How to strike a proper balance between pre-trained and newly-trained components remains an open question.


Pre-training the model on MS-COCO followed by fine-tuning on M+T+V brings a marginal improvement, \ie from $0.240$ (Rank 2) to $0.249$ (Rank 1). The relatively lower performance of rank 3, with a mean infAP of $0.230$, shows that GCC is less effective than MS-COCO for pre-training.


\begin{table}[]
\normalsize
\renewcommand\arraystretch{1}
\centering

\begin{center}
\addvbuffer[0pt 0pt]{
\scalebox{0.81}{
\begin{tabular}{|l|r|r|r|r|r|r|}

\hline
\textbf{id} & \multicolumn{1}{l|}{\textbf{Rank1}} & \multicolumn{1}{l|}{\textbf{Rank2}} & \multicolumn{1}{l|}{\textbf{Rank4}} & \multicolumn{1}{l|}{\textbf{Rank12}} & \multicolumn{1}{l|}{\textbf{Rank18}} & \multicolumn{1}{l|}{\textbf{MEAN}} \\ \hline
644 & \cellcolor[HTML]{F8696B}0.852 & \cellcolor[HTML]{F96C6E}0.843 & \cellcolor[HTML]{82A6D4}0.067 & \cellcolor[HTML]{F98D8F}0.700 & \cellcolor[HTML]{FAA1A4}0.614 & \cellcolor[HTML]{FAA1A3}0.615 \\ \hline
642 & \cellcolor[HTML]{FA989A}0.653 & \cellcolor[HTML]{F98D8F}0.700 & \cellcolor[HTML]{FAA1A3}0.616 & \cellcolor[HTML]{FCE8EA}0.314 & \cellcolor[HTML]{FCF6F9}0.251 & \cellcolor[HTML]{FBBABD}0.507 \\ \hline
656 & \cellcolor[HTML]{FA9C9E}0.636 & \cellcolor[HTML]{FA9597}0.666 & \cellcolor[HTML]{FA9FA1}0.625 & \cellcolor[HTML]{FCDFE2}0.351 & \cellcolor[HTML]{EBF0F9}0.204 & \cellcolor[HTML]{FBBDC0}0.496 \\ \hline
660 & \cellcolor[HTML]{FBB7B9}0.522 & \cellcolor[HTML]{FBC4C7}0.464 & \cellcolor[HTML]{FBC0C3}0.483 & \cellcolor[HTML]{FBB6B8}0.526 & \cellcolor[HTML]{FCF7FA}0.248 & \cellcolor[HTML]{FBC8CB}0.449 \\ \hline
649 & \cellcolor[HTML]{FA9193}0.684 & \cellcolor[HTML]{FAA6A9}0.593 & \cellcolor[HTML]{FBCCCF}0.431 & \cellcolor[HTML]{FCE2E5}0.337 & \cellcolor[HTML]{98B5DB}0.095 & \cellcolor[HTML]{FBCDD0}0.428 \\ \hline
651 & \cellcolor[HTML]{FA9799}0.658 & \cellcolor[HTML]{FAA0A3}0.618 & \cellcolor[HTML]{FCDDE0}0.358 & \cellcolor[HTML]{D9E3F2}0.180 & \cellcolor[HTML]{F4F6FC}0.215 & \cellcolor[HTML]{FBD2D5}0.406 \\ \hline
659 & \cellcolor[HTML]{FBD2D5}0.407 & \cellcolor[HTML]{FBCDD0}0.428 & \cellcolor[HTML]{FBD3D6}0.401 & \cellcolor[HTML]{FCEDEF}0.293 & \cellcolor[HTML]{FCFBFE}0.233 & \cellcolor[HTML]{FCDFE2}0.352 \\ \hline
654 & \cellcolor[HTML]{FCDEE1}0.355 & \cellcolor[HTML]{FBD7DA}0.385 & \cellcolor[HTML]{FCDBDE}0.366 & \cellcolor[HTML]{FCE4E7}0.330 & \cellcolor[HTML]{DDE6F4}0.185 & \cellcolor[HTML]{FCE5E8}0.324 \\ \hline
647 & \cellcolor[HTML]{FCE8EB}0.312 & \cellcolor[HTML]{FCEEF1}0.287 & \cellcolor[HTML]{FBCBCD}0.437 & \cellcolor[HTML]{F1F4FB}0.211 & \cellcolor[HTML]{DAE4F3}0.181 & \cellcolor[HTML]{FCEEF1}0.286 \\ \hline
653 & \cellcolor[HTML]{FCD9DC}0.376 & \cellcolor[HTML]{FCDCDF}0.364 & \cellcolor[HTML]{FCDDE0}0.358 & \cellcolor[HTML]{85A8D5}0.071 & \cellcolor[HTML]{719ACE}0.045 & \cellcolor[HTML]{FCF8FB}0.243 \\ \hline
641 & \cellcolor[HTML]{FCF0F3}0.280 & \cellcolor[HTML]{FCF7FA}0.250 & \cellcolor[HTML]{FCEEF1}0.288 & \cellcolor[HTML]{C7D6EC}0.156 & \cellcolor[HTML]{E6ECF7}0.197 & \cellcolor[HTML]{FCFAFD}0.234 \\ \hline
652 & \cellcolor[HTML]{FCFBFE}0.233 & \cellcolor[HTML]{FCFAFD}0.236 & \cellcolor[HTML]{FCFBFE}0.231 & \cellcolor[HTML]{F7F8FD}0.219 & \cellcolor[HTML]{FCF7FA}0.250 & \cellcolor[HTML]{FCFAFD}0.234 \\ \hline
646 & \cellcolor[HTML]{E4EBF6}0.195 & \cellcolor[HTML]{EAEFF8}0.202 & \cellcolor[HTML]{FCF5F7}0.259 & \cellcolor[HTML]{F4F6FC}0.215 & \cellcolor[HTML]{7BA1D1}0.058 & \cellcolor[HTML]{DEE6F4}0.186 \\ \hline
643 & \cellcolor[HTML]{DBE5F3}0.183 & \cellcolor[HTML]{C6D6EC}0.155 & \cellcolor[HTML]{CFDCEF}0.167 & \cellcolor[HTML]{9FBADE}0.104 & \cellcolor[HTML]{A5BEE0}0.112 & \cellcolor[HTML]{BDD0E9}0.144 \\ \hline
645 & \cellcolor[HTML]{8CADD7}0.080 & \cellcolor[HTML]{8EAED8}0.082 & \cellcolor[HTML]{AAC2E2}0.119 & \cellcolor[HTML]{7DA2D2}0.060 & \cellcolor[HTML]{A6C0E1}0.114 & \cellcolor[HTML]{95B3DA}0.091 \\ \hline
657 & \cellcolor[HTML]{95B3DA}0.091 & \cellcolor[HTML]{84A7D4}0.069 & \cellcolor[HTML]{A6C0E1}0.114 & \cellcolor[HTML]{82A6D4}0.067 & \cellcolor[HTML]{84A8D5}0.070 & \cellcolor[HTML]{8EAED8}0.082 \\ \hline
650 & \cellcolor[HTML]{96B4DB}0.093 & \cellcolor[HTML]{88AAD6}0.075 & \cellcolor[HTML]{A8C1E1}0.116 & \cellcolor[HTML]{6994CB}0.034 & \cellcolor[HTML]{5C8BC6}0.017 & \cellcolor[HTML]{82A6D4}0.067 \\ \hline
648 & \cellcolor[HTML]{83A7D4}0.068 & \cellcolor[HTML]{87A9D5}0.073 & \cellcolor[HTML]{99B6DC}0.097 & \cellcolor[HTML]{7DA2D2}0.060 & \cellcolor[HTML]{6592CA}0.029 & \cellcolor[HTML]{81A5D3}0.065 \\ \hline
655 & \cellcolor[HTML]{7AA0D1}0.056 & \cellcolor[HTML]{82A6D4}0.067 & \cellcolor[HTML]{8BACD7}0.078 & \cellcolor[HTML]{5A8AC6}0.014 & \cellcolor[HTML]{789FD0}0.054 & \cellcolor[HTML]{789FD0}0.054 \\ \hline
658 & \cellcolor[HTML]{6F99CD}0.042 & \cellcolor[HTML]{628FC8}0.025 & \cellcolor[HTML]{81A6D4}0.066 & \cellcolor[HTML]{6592CA}0.029 & \cellcolor[HTML]{6C96CC}0.038 & \cellcolor[HTML]{6D98CD}0.040 \\ \hline
\end{tabular}
}}
    
\end{center}

\caption{\textbf{Per-query comparison on TV20}. Five models (Rank 1 / 2 / 4 / 12 / 18) are selected from Table \ref{tab:infAP_of_each_model_ranked}, which differ in terms of matching models, visual features, and (pre-)training data. Queries are sorted by the mean performance of the five  models. }
\label{tab:per-topic_analysis}

\end{table}
\section{Discussion and Concluding Remarks}







Towards answering the question of what matters for ad-hoc video search, we have presented a large-scale empirical study on the TRECVID AVS benchmarks in the last five years (2016--2020). When using \emph{clip} as a single visual feature, the lightweight CLIP-FRL model outperforms the six other models. This can be attributed to the joint effect of the pre-trained CLIP and the newly added FC layer trained with respect to the video-text matching task. Probably due to the short nature of the videos used in the AVS task, the deep 3D features, \ie \emph{c3d}, \emph{ircsn}, and \emph{tf}, are less effective than their 2D alternatives when used alone. Nonetheless, they remain helpful when used in combination with the 2D features. Adding the VATEX dataset into the training data brings in noticeable performance gain in particular on TV17 and TV20, as this dataset contains a number of novel query words which are either absent or rarely occur in the frequently used MSR-VTT and TGIF. MS-COCO, for its high-quality annotations, is better than GCC for pretraining.

It is worth pointing out that as the number of the AVS test queries is relatively small, the performance of a specific solution can be affected by few queries that perform either extremely well or extremely bad. We consider it necessary to enlarge the test query set for a more stable evaluation.

\medskip
\textbf{Acknowledgements}. This research was supported by NSFC (No. 61672523), BJNSF (No. 4202033), the Fundamental Research Funds for the Central Universities and the Research Funds of Renmin University of China (No. 18XNLG19).

{\small
\bibliographystyle{ieee_fullname}
\bibliography{egpaper_final}
}

\end{document}